\documentclass[11pt,a4paper]{article}
\pdfoutput=1

\usepackage{jheppub} 

\usepackage{hyperref,graphicx,color,amsmath,amssymb,slashed,xspace,rotating}

\newcommand{\GeV}{\,\text{GeV}}
\newcommand{\TeV}{\,\text{TeV}}

\newcommand{\fastjet}{\textsf{FastJet}\xspace}
\newcommand{\pythia}{\textsf{Pythia}}
\newcommand{\rivet}{\textsf{Rivet}\xspace}

\title{Vetoed jet clustering: The mass-jump algorithm}
\author{Martin Stoll}
\vspace{5mm}
\affiliation{\it Department of Physics, University of Tokyo, Bunkyo-ku, Tokyo 113--0033, Japan}
\vspace{1cm}
\emailAdd{stoll@hep-th.phys.s.u-tokyo.ac.jp}
\abstract{A new class of jet clustering algorithms is introduced. A criterion inspired by successful mass-drop taggers is applied that prevents the recombination of two hard prongs if their combined jet mass is substantially larger than the masses of the separate prongs. This ``mass jump'' veto effectively results in jets with variable radii in dense environments.
Differences to existing methods are investigated. It is shown for boosted top quarks that the new algorithm has beneficial properties which can lead to improved tagging purity.
}
\preprint{UT--14--41}

\begin{document}
\maketitle
\flushbottom


\section{Introduction}
\label{sec:introduction}

  The Large Hadron Collider (LHC) will restart in 2015 with an unprecedented centre-of-mass energy, 
  offering a new opportunity to discover yet unknown particles beyond the Standard Model (SM).
  Practically all processes --- SM or hypothetical --- contain quarks or gluons in the final state and it is important that they can be reconstructed reliably.
  These coloured partons undergo parton showering and hadronization before they leave a signal in the detector. In order to make sense of experimental data it is therefore necessary to collect nearby radiation into jets, which are then assumed to correspond to their initiating (hard) partons.
  
  Whenever jets are used as input for an analysis, the significance of the results crucially depends on the validity of this kinematic correspondance. Hence there has been ongoing effort to construct new and improved jet algorithms that are infrared and collinear safe, most of which proceed via sequential recombination~\cite{Catani:1991hj,Catani:1993hr,Ellis:1993tq, Dokshitzer:1997in,Wobisch:1998wt, Cacciari:2008gp} or cones~\cite{Sterman:1977wj, Blazey:2000qt, Albrow:2006rt, Salam:2007xv}, or follow completely different original ideas~\cite{Bertolini:2013iqa, Georgi:2014zwa,Ge:2014ova}.
  In the majority of these algorithms, jets are constructed with fixed angular size $R$, defined between two particles as $\Delta R=\sqrt{\Delta y^2+\Delta\phi^2}$ where $\Delta y$ and $\Delta\phi$ are the distances in rapidity and azimuthal angle, respectively.
  
  Despite this splendour of algorithms to select from, choosing the optimal jet radius is always a compromise~\cite{Dasgupta:2007wa,Cacciari:2008gd,Soyez:2010rg} as it may be different for jets of different energy or position in the detector. Ref.~\cite{Krohn:2009zg} consequently proposes to employ a variable clustering radius instead, which in this case is taken inversely proportial to the jet transverse momentum, $R \propto 1/p_\perp$.
  An entirely different approach is taken by mass-drop tagging algorithms~\cite{Butterworth:2008iy, Plehn:2010st, Schaetzel:2013vka}. They address heavy resonances that are so highly boosted that their subsequent decay products cannot reasonably be resolved with conventional jet algorithms.
  Due to the high centre-of-mass energy of the LHC, boosted top quarks, Higgs bosons, etc.~are expected to be produced in larger numbers during the upcoming run.
  To identify these resonances, it is possible to capture all decay products in a large-radius fat jet and apply substructure methods. The basic idea states that a jet should be broken up into two separate subjets if the jet mass experiences a significant drop in the procedure. 
  These algorithms identify hard substructure without referring to a fixed (sub)jet radius and turned out to perform very well in Higgs boson and top quark tagging (see e.g.~refs.~\cite{Abdesselam:2010pt,Altheimer:2012mn,Plehn:2011tg} for reviews).
  Implicitly, a $p_\perp$-dependent subjet radius is given by the mass cut, as the characteristic separation between the daughters of an energetic resonance is $\Delta R_\text{daughters}\gtrsim 2m_\text{mother}/p_\perp$.
  
  In this paper, we supplement existing jet algorithms with a recombination veto, which may prevent further clustering at a jet radius smaller than the given $R$. The working principle is similar to mass-drop tagging: if the recombination of two jet candidates leads to a significant mass jump, they should be resolved separately.
  In contrast to algorithms with variable radius, the veto is a property of two jets, i.e.~the effective clustering radius now also depends on the jet's vicinity.
  This way well-separated jets are clustered conventionally with only small deviations, whereas on the other hand the merging of two hard prongs into a heavy resonance is vetoed.
  
  The introduction of a clustering veto is not a novelty. One notable example is given by pruning~\cite{Ellis:2009su,Ellis:2009me}, which however follows a different philosophy. Here a recombination step is vetoed if it resembles large-angle soft radiation (expressed in terms of transverse momentum and $R$ separation) in the sense that jet clustering proceeds as usual after the softer part has been discarded from the event. This way only hard substructure is kept and the algorithm can already be used as a tagger. Algorithms which remove soft uncorrelated radiation (from underlying event or pile-up) are collectively called groomers (ref.~\cite{Plehn:2011tg} gives a brief and comprehensive overview of the most common algorithms filtering~\cite{Butterworth:2008iy}, trimming~\cite{Krohn:2009th}, and the beforementioned pruning).
  In contrast, here we suggest a terminating veto for the mass-jump procedure: when the merging of two hard prongs is vetoed, they no longer participate in jet clustering. This way (sub)jets are identified without reference to an external energy or angular scale, while keeping all the radiation present in the event.
  
  This paper is organized as follows. In section~\ref{sec:algorithm}, the mass-jump algorithm is motivated and described in detail. Throughout the paper, we focus on consequences of the recombination veto in comparison to both classic jet algorithms as well as mass-drop taggers. In section~\ref{sec:performance}, we first evaluate the peformance for well-separated jets, and then turn to the boosted regime. Beneficial properties for top quark tagging are pointed out. Conclusions are drawn in section~\ref{sec:conclusions}.

\section{The algorithm}
\label{sec:algorithm}

\subsection{Review: mass-drop unclustering}
\label{sec:algorithm:md}

  Developed to identify boosted Higgs bosons decaying into a pair of bottom quarks, the BDRS Higgs Tagger~\cite{Butterworth:2008iy} established the family of mass-drop tagging (MDT) algorithms. 
  The goal of this algorithm is to identify the 2-prong substructure of the decay $H\to b\bar{b}$ within one wide-angle (``fat'') jet.
  The modified 3-prong variant, the HEPTopTagger~\cite{Plehn:2010st}, enforces the following iterative procedure to act on a given fat jet clustered with the Cambridge/Aachen jet algorithm~\cite{Dokshitzer:1997in,Wobisch:1998wt}.
  
  \begin{itemize}
   \item Undo the last clustering of the jet $j$ into $j_1$, $j_2$, ordered $m_{j_1}>m_{j_2}$.
   \item If a significant mass drop occurred, $m_{j_1}<\theta\cdot m_j$, both $j_1$ and $j_2$ are kept as candidate subjets. Otherwise discard $j_2$.\footnote{It has been pointed out in a related setup~\cite{Dasgupta:2013ihk} that following the heavier prong leads to a (small) wrong-branch contribution. This can be avoided by discarding the subjet candidate with smaller transverse mass $m_\perp^2\equiv m^2+p_\perp^2$ instead. As this modification is irrelevant for the remainder of this paper, we do not distinguish between the MDT and this modified mass-drop tagger (mMDT).}
   \item Repeat these steps for the kept subjets unless $m_{j_i} < \mu$, in which case $j_i$ is added to the set of output subjets.
  \end{itemize}
  
  The mass-drop (MD) procedure\footnote{Note that in the literature, sometimes the expressions ``mass drop'' and ``mass drop tagger'' are used to explicitly refer to the original BDRS Higgs tagging algorithm~\cite{Butterworth:2008iy}. There, the mass-drop condition is supplemented with a symmetry criterion $y=\min(p_{\perp,j_1}^2,p_{\perp,j_2}^2)/m_j^2 \cdot\Delta R_{j_1,j_2}^2 > y_\text{cut}$ motivated by the decay $H\to b\bar{b}$. Analytic calculations for isolated jets have shown~\cite{Dasgupta:2013ihk} that in this algorithm the dependence on the mass-drop parameter $\theta$ is actually only small. This observation is used for the soft drop procedure~\cite{Larkoski:2014wba}, which is solely defined in terms of a generalized symmetry criterion, $\min(p_{\perp,j_1},p_{\perp,j_2})/p_{\perp,j} > z_\text{cut} (\Delta R_{j_1,j_2}/R_0)^\beta$ (with parameters $R_0$ and $\beta$), and is interesting in its own respect.
  In this paper, however, we focus on the plain mass-drop condition as defined in the text and implemented in the HEPTopTagger~\cite{Plehn:2010st}. It is expected to be better suited for general decay patterns or event kinematics and has been proven very successful for top quark tagging (see e.g.~ref.~\cite{Altheimer:2012mn}). Below we will develop a new algorithm based on this reading of ``mass drop''.} 
  serves two purposes: It grooms the jet from (large-angle) soft radiation and applies a criterion to identify a non-specified number of separate prongs based on jet mass.
  In the HEPTopTagger, the set of output subjets is then further processed and cuts applied. The default values of the two free parameters are chosen as $\theta=0.8$ and $\mu=30\GeV$~\cite{Plehn:2010st,Anders:2013oga}.
  
  Note that the un-clustering algorithm is designed to follow the cascade decay chain of the top quark,
  \begin{align}
   t \to b W^+ \to b j j' \,.
  \end{align}
  At parton level the successive mass drops $\tau=\frac{m_{j_1}}{m_j}$ are given by
  \begin{align}
   \tau_1 = \frac{m_W}{m_t} \approx 0.46 \,,\quad \tau_2 = \frac{m_q}{m_W} \approx 0 \,,
  \end{align}
  hence the parameter $\theta$ has to be chosen sufficiently large to incorporate the first decay. In case the unclustering proceeds via $t \to j' (bj) \to j' b j$ one obtains
  \begin{align}
   \tau_1' = \frac{\sqrt{m_t^2-m_W^2}}{2m_t} \Delta R_{bj} \,,
  \end{align}
  which is typically smaller than $\tau_1$. $\Delta R_{bj} = \sqrt{\Delta y^2 +\Delta\phi^2}$ is the $R$-distance between the subjets $b$ and $j$.
  
\subsection{The mass-jump clustering algorithm}

  Commonly used sequential jet clustering algorithms define an infrared and collinearly safe procedure to merge particles into jets step by step. Termination of this sequential recombination is given (in the inclusive algorithms) in terms of a minimum jet separation $R$. All input particles 
  are labelled as jet candidates and a distance measure betweens pairs of two is defined,
  \begin{align}
   d_{j_1j_2} = \frac{\Delta R_{j_1j_2}^2}{R^2} \min\left[p_{j_1\perp}^{2n},p_{j_2\perp}^{2n}\right] \,,\quad
   d_{j_1B} = p_{j_1\perp}^{2n} \,,
  \end{align}
  where $n=1$ corresponds to the $k_T$ algorithm~\cite{Catani:1991hj,Catani:1993hr,Ellis:1993tq}, $n=0$ to the Cambridge/Aachen algorithm~\cite{Dokshitzer:1997in,Wobisch:1998wt}, and $n=-1$ to the anti-$k_T$ algorithm~\cite{Cacciari:2008gp}.
  Sequential recombination then proceeds as follows:
  \begin{enumerate}
   \item Find the smallest $d_{j_aj_b}$ among the jet candidates.
   If it is given by a beam distance, $d_{j_aB}$, label $j_a$ a jet and repeat step 1.
   \item Otherwise combine $j_a$ and $j_b$ by summing their four-momenta, $p_{j_aj_b}=p_{j_a}+p_{j_b}$ ($E$-scheme, see e.g.~ref.~\cite{Blazey:2000qt}). In the set of jet candidates, replace $j_a$ and $j_b$ by their combination and go back to step 1.
  \end{enumerate}
  Clustering eventually ends when all particles have been merged into jets.
  The measure $d$ serves two purposes here: first, it determines the order of recombination given by the pair with the smallest distance $d_{j_aj_b}$ at each step. Second, it acts as an upper bound on the jet radius, because a minimal beam distance $d_{j_aB}$ implies $\Delta R_{j_aj_n}>R$ $\forall$ jet candidates $j_n$.

  We present a modification to these jet clustering algorithms which we call mass-jump (MJ) clustering. In the spirit of a reverse mass-{\it drop} procedure as outlined in the previous paragraph, ``sub''jets are directly constructed by examining a veto condition at each recombination step,\footnote{Separate measures for ordering variable and test (veto) variable were first introduced in ref.~\cite{Dokshitzer:1997in}.}
  where the parameter $\theta$ now acts as a mass-{\it jump} threshold.
  After all input particles 
  are labelled as {\it active} jet candidates, the recombination algorithm is defined as follows:
  \begin{enumerate}
   \item Find the smallest $d_{j_aj_b}$ among active jet candidates; if it is given by a beam distance, $d_{j_aB}$, label $j_a$ {\it passive} and repeat step 1.
   \item Combine $j_a$ and $j_b$ by summing their four-momenta, $p_{j_a+j_b}=p_{j_a}+p_{j_b}$ ($E$-scheme).
   If the new jet is still light, $m_{j_a+j_b}<\mu$, replace $j_a$ and $j_b$ by their combination in the set of active jet candidates and go back to step 1.\\
   Otherwise check the mass-jump criterion: if $\theta\cdot m_{j_a+j_b} >  \max\left[m_{j_a},m_{j_b}\right]$ label $j_a$ and $j_b$ {\it passive} and go back to step 1.
   
   \item Mass jumps can also appear between an active and a passive jet candidate.
   To examine this
   \begin{itemize}
    \item[a.] Find the passive jet candidate $j_n$ that is closest to $j_a$ in terms of the metric $d$ and is not isolated, $d_{j_aj_n}<d_{j_nB}$.
    \item[b.] Then check if these two jet candidates would have been recombined if $j_n$ had not been rendered passive by a previous veto, i.e.~$d_{j_aj_n}<d_{j_aj_b}$.
    \item[c.] Finally check the mass-jump criterion, $m_{j_a+j_n}\geq\mu$ and $\theta\cdot m_{j_a+j_n}>\max\left[m_{j_a},m_{j_n}\right]$.
   \end{itemize}
   If all these criteria for the veto are fulfilled, label $j_a$ {\it passive}.
   Do the same for $j_b$.
   If either of $j_a$ or $j_b$ turned passive, go back to step 1.
   
   
   \item No mass jump has been found, so replace $j_a$ and $j_b$ by their combination in the set of active jet candidates. Go back to step 1.
  \end{enumerate}
  Clustering terminates when there are no more active jet candidates left. Passive candidates are then labelled jets.
  Note that for $\theta=0$ or $\mu=\infty$ this algorithm is identical to standard sequential clustering without veto, which in this case can be reduced to steps 1 and 4.

\subsection{Properties}
\label{sec:algorithm:properties}

  The mass-jump veto only has an impact on jet candidates that are separated by $\Delta R<R$ and whose combined mass would be above the (arbitrary) scale $\mu$. It is designed to resolve close-by jets (which could come from the decay of a boosted resonance such as $W^\pm$, $Z$, $H$, ...) separately. As the vetoed jets are excluded from further clustering, their effective jet radius is smaller than the parameter $R$, which now gives an upper bound. A lower bound is indirectly induced by a finite threshold scale $\mu$.
  
\begin{figure}
   \begin{center}
   mass-drop unclustering of a fat jet\\
   \includegraphics[trim=3cm 14cm 2cm 2cm, clip=true, totalheight=0.32\textheight]{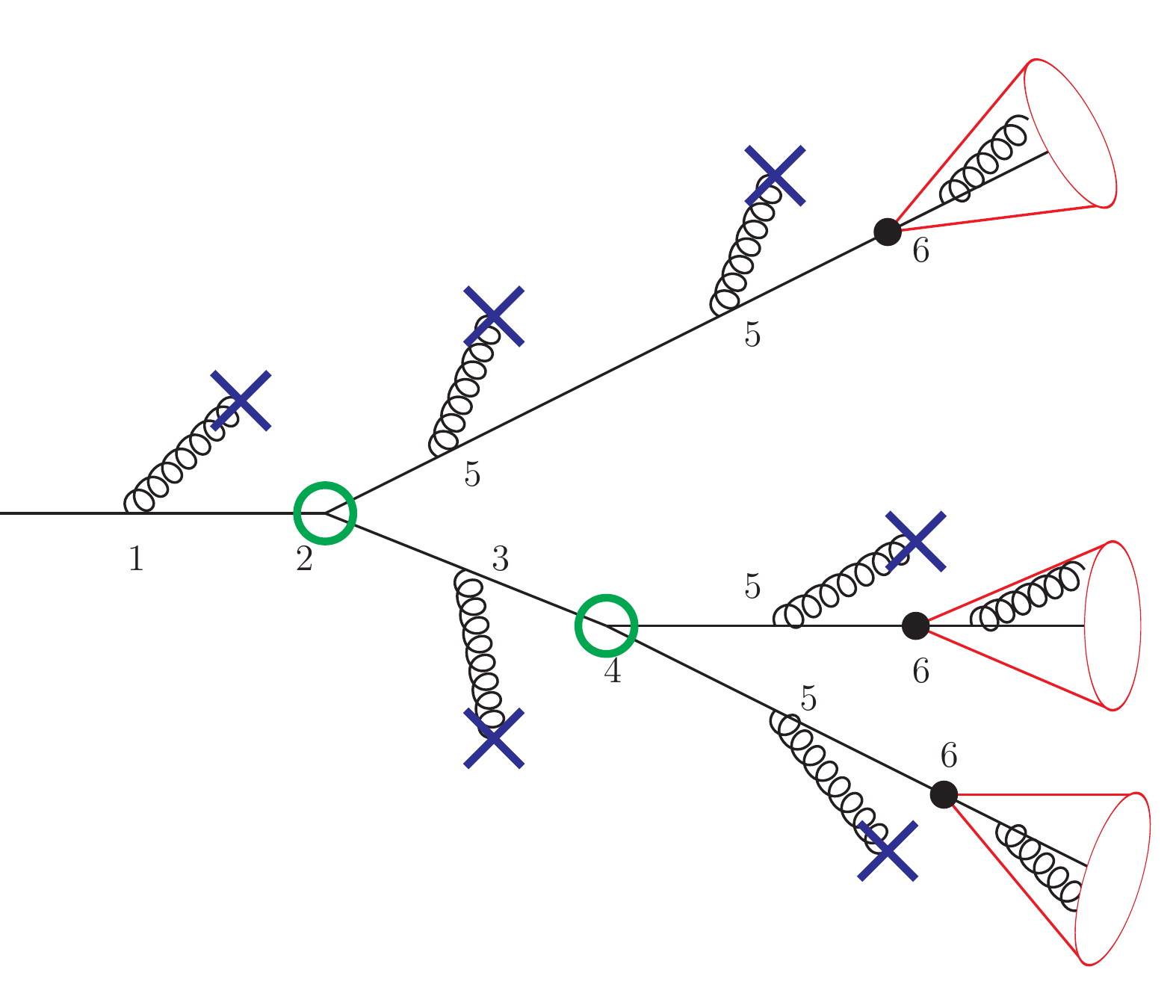}
   
   mass-jump clustering\\
   \includegraphics[trim=3cm 14cm 2cm 2cm, clip=true, totalheight=0.32\textheight]{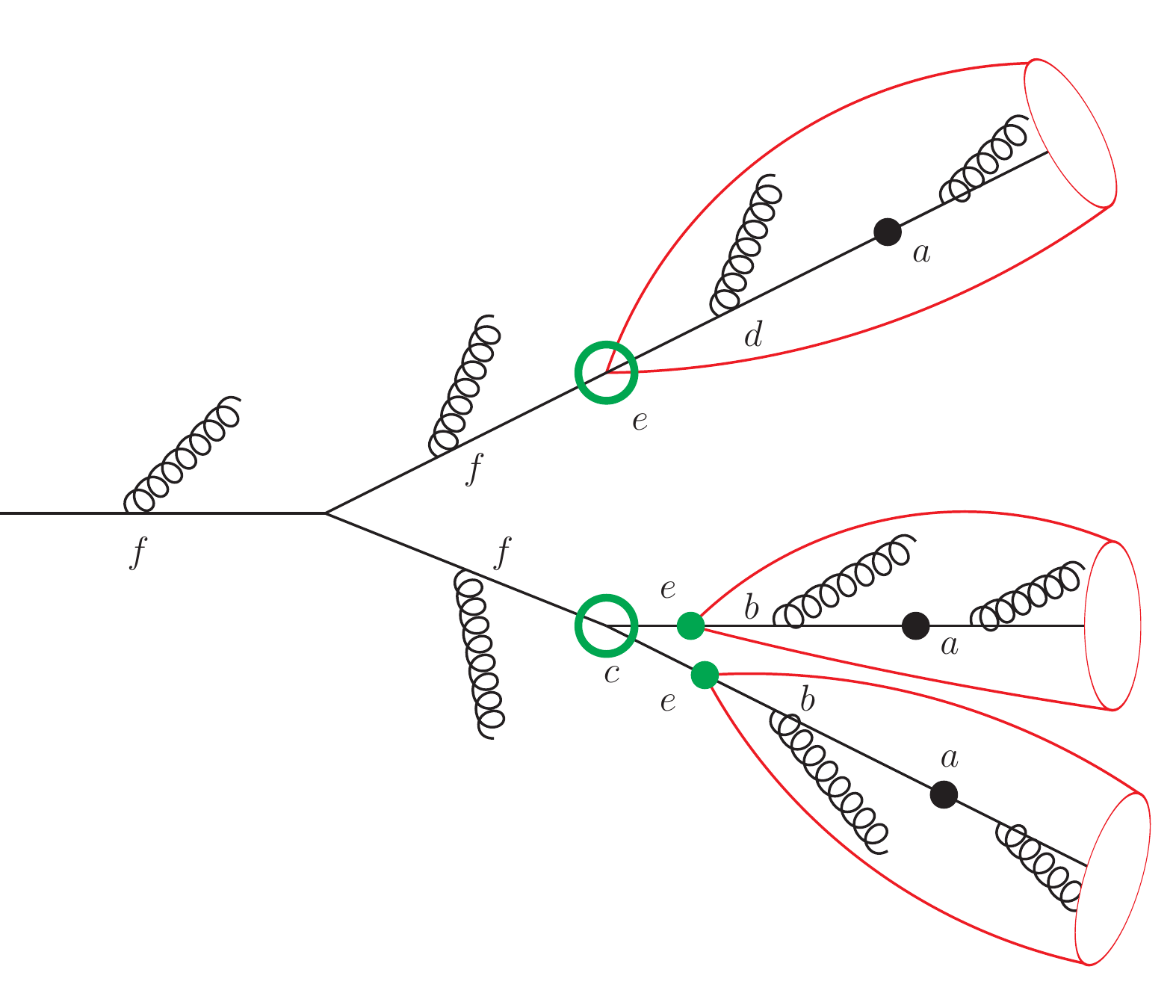}
   
   \caption{Key differences between MD unclustering (top) and MJ clustering (bottom) are visualized for a schematic clustering sequence (e.g.~of a hadronically decaying boosted top quark). Sequential recombination is performed starting from the constituents at the right-hand side, such that in the upper panel the very left line symbolizes the whole fat jet, which is then sequentially unclustered again (bottom panel: MJ clustering works its way from the constituent particles to the left). Inside the cluster sequence, hard prongs are depicted as straight lines, whereas wiggly lines symbolize soft radiation. Black dots denote the jet mass threshold $m=\mu$, and green circles indicate a mass drop (or mass jump). The final (sub)jets are indicated with red cones. The individual steps of the respective two algorithms (steps 1--6 for MD unclustering, steps a--f for MJ clustering) are described in the text.}
   \label{fig:cluster-sequence}
   \end{center}
  \end{figure}

  There are several similarities and differences compared to MD unclustering. Figure~\ref{fig:cluster-sequence} schematically depicts a standard clustering sequence (e.g.~of a hadronically decaying boosted top quark) and how the two algorithms act on the given event. The clustering sequence is to be read from right to left; hard prongs are depicted as straight lines, whereas wiggly lines symbolize soft radiation. The MDT sequentially unclusters a fat jet (which can be an actual large-radius jet or the whole event) from left to right, whereas the MJ algorithm starts from the fat jet's constituents and proceeds to the left. The final (sub)jets are indicated by red cones.
  
  In the MDT algorithm (upper panel), starting from a fat jet soft radiation is groomed away (1) until at one unclustering step the mass-drop criterion is fulfilled, resulting in two subjets (2). The same grooming--tagging procedure continues for every prong that experiences a further mass drop (3+4). More soft radiation is removed (5) until the subjet masses are below the threshold $\mu$ (6). The remaining prongs are now labelled ``subjets''.
  
  MJ clustering (lower panel), on the other hand, is identical to standard clustering algorithms until the jet mass exceeds $\mu$ (a).\footnote{Or the jet has reached its size given by the radius $R$ -- for the sake of comparison with the MDT procedure, we take $R=\infty$ for the moment.}
  Clustering continues (b) until the next recombination step would result in a substantial mass jump (c), at which step clustering is vetoed and the two prongs turn passive. Active jet candidates continue clustering (d) unless a veto is called, which can also act against a (hypothetical) recombination with a passive jet (e). Jet clustering continues for the remaining particles, giving additional jets (f).
  
  In the idealized case, the output jets of both algorithms are comparable but differ in two aspects.
  First, MDT subjets are groomed even after a mass drop until they reach $m<\mu$ whereas MJ jets continue collecting radiation in the regime between $m>\mu$ and the mass jump. Although this effect is expected to be absent for reasonably large values of $\mu$, if undesired it is straightforward to apply MDT-like grooming on the MJ jets.
  Second, the MJ clustering algorithm also returns jets that did not experience mass jumps (f) that are absent among MDT subjets (1,3,5). These can be desirable (well-separated jets for finite $R$) or can be considered junk; in the latter case it is again straightforward to remove them as these are the only jets turned {\it passive} by the upper bound on the jet radius instead of a mass jump.
  
  Also note the important property that MD unclustering experiences cascade mass drops (cf.~section~\ref{sec:algorithm:md}) while MJ clustering does not. This results in all mass jumps being among single hard prongs with a typical scale $\sim {m_\text{heavy resonance}}/{\mu}$, i.e.~the threshold parameter $\theta$ can be chosen substantially lower.

\section{Performance}
\label{sec:performance}

\subsection{Sparse environment: QCD dijets}

  We compare the MJ clustering algorithm to its standard counterparts. 
  QCD dijet events are expected to contain two well-separated hard jets, however more jets may be found due to large-angle emissions or jet substructure induced by the parton shower.
  In particular MJ clustering is prone to misidentify jet substructure as separate hard objects, and this section aims to quantify this effect of the veto.
  10,000 QCD dijet events are simulated with \pythia8~\cite{Sjostrand:2007gs} where the minimum parton transverse momentum at matrix element level is chosen $\hat{p}_\perp^\text{min}=40\GeV$.
  The analysis is implemented as a \rivet~\cite{Buckley:2010ar} plugin.
  
  Jets are constructed from all (visible) final-state particles with pseudo-rapidity $|\eta|\leq 4.9$.
  The clustering parameters are chosen $R=0.8$ and $p_\perp\geq p_\perp^\text{min}=50\GeV$, also jets are required to be sufficiently central, $|\eta|\leq 4.0$.
  We compare the jets clustered with a standard algorithm (anti-$k_T$, Cambridge/Aachen, or $k_T$ algorithms as provided by \fastjet~\cite{Cacciari:2011ma})
  to those obtained with the corresponding MJ algorithm on an event-by-event basis.  
  Only events that contain at least one hard jet from the standard algorithm, $p_\perp^{\text{std}(1)}\geq 150\GeV$, are accepted.
  This assures that the leading jet is still present among the MJ jets and does not drop below $p_\perp^\text{min}$, even if torn apart by the clustering veto.
  For each of the algorithms, $\sim 100$ events pass this cut.
  
\begin{figure}[t]
  \begin{center}
   \begin{minipage}{0.02\textwidth}
   \end{minipage}
   \begin{minipage}{0.31\textwidth}
    \centering
    anti-$k_T$
   \end{minipage}
   \begin{minipage}{0.31\textwidth}
    \centering
    C/A
   \end{minipage}
   \begin{minipage}{0.31\textwidth}
    \centering
    $k_T$
   \end{minipage}

    \begin{minipage}[t]{0.021\textwidth}
    \begin{sideways}$\qquad\qquad \Delta R$\end{sideways}
    \end{minipage}
    \includegraphics[trim=0cm 0cm 1.7cm 0cm, clip=true, height=0.23\textwidth]{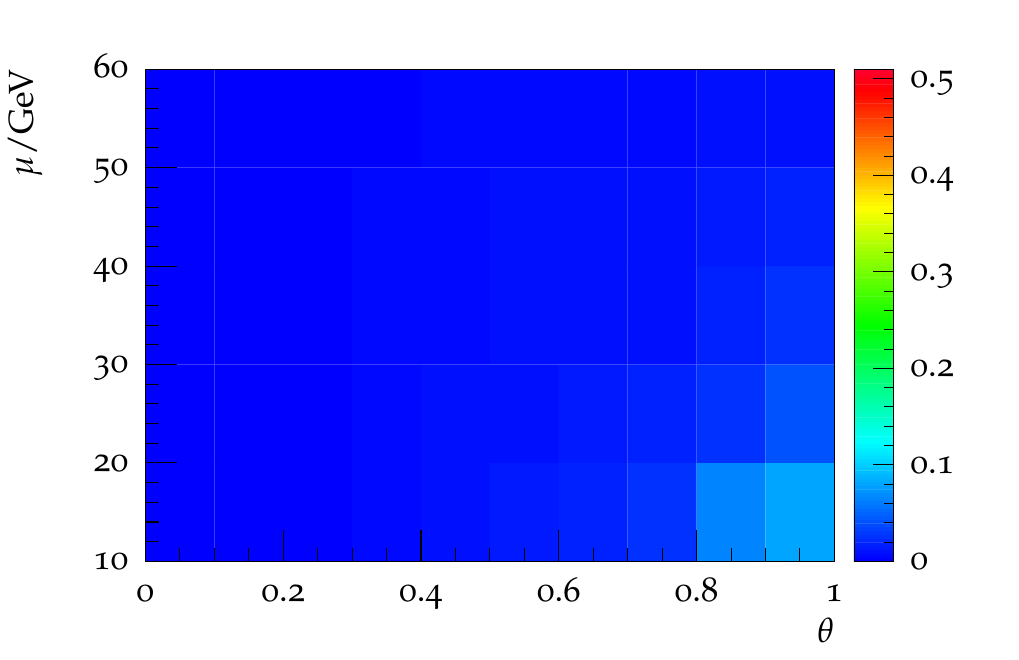}
    \includegraphics[trim=0cm 0cm 1.7cm 0cm, clip=true, height=0.23\textwidth]{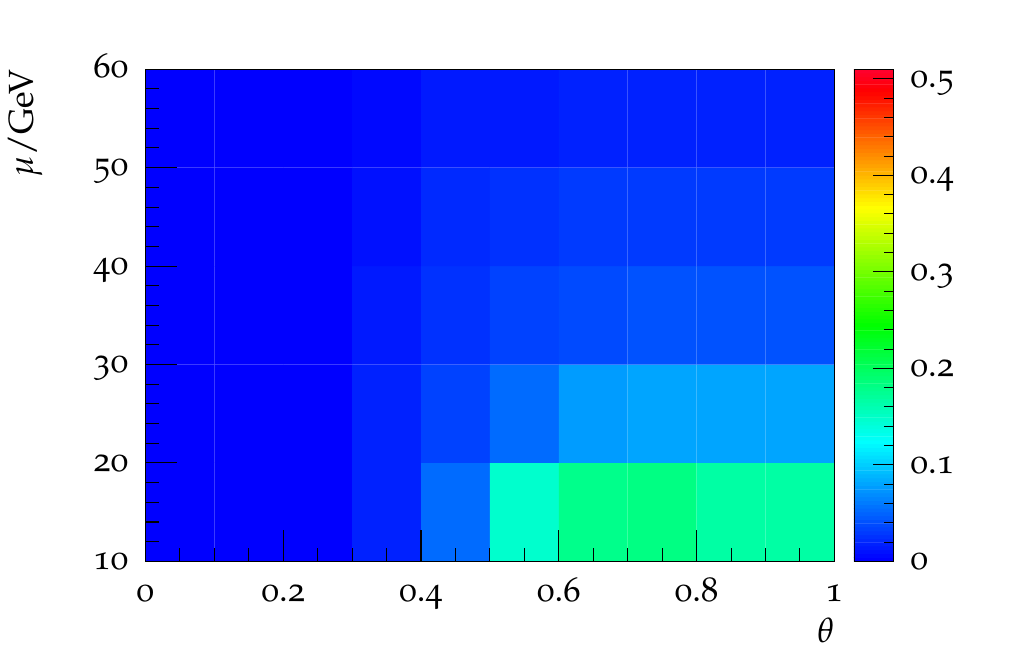}
    \includegraphics[height=0.23\textwidth]{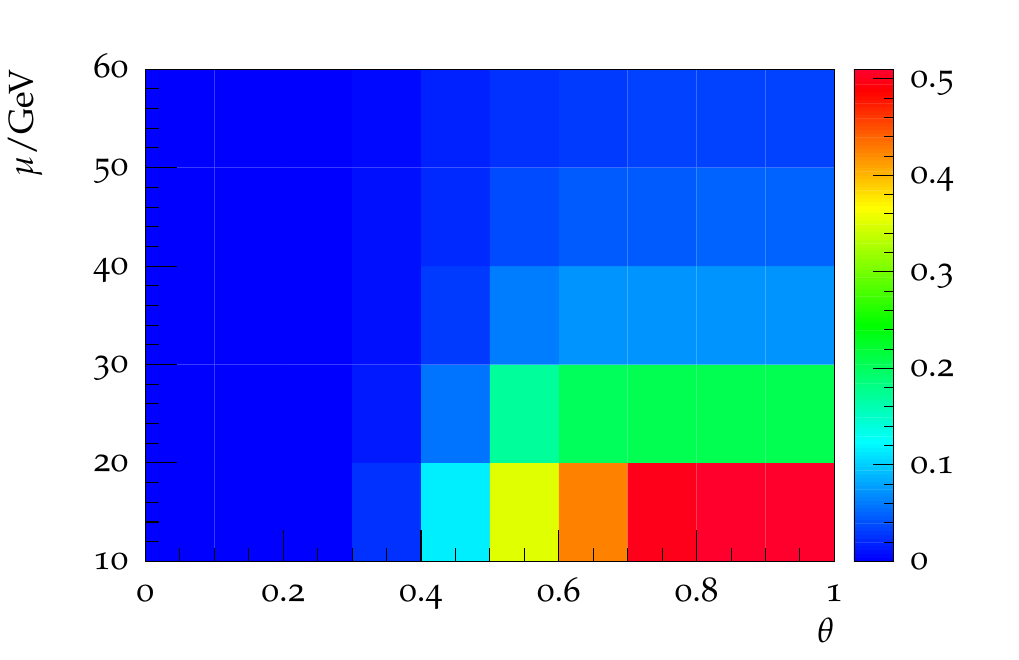}
    
    \begin{minipage}[t]{0.021\textwidth}
    \begin{sideways}$\qquad\qquad \delta p_\perp$\end{sideways}
    \end{minipage}
    \includegraphics[trim=0cm 0cm 1.7cm 0cm, clip=true, height=0.23\textwidth]{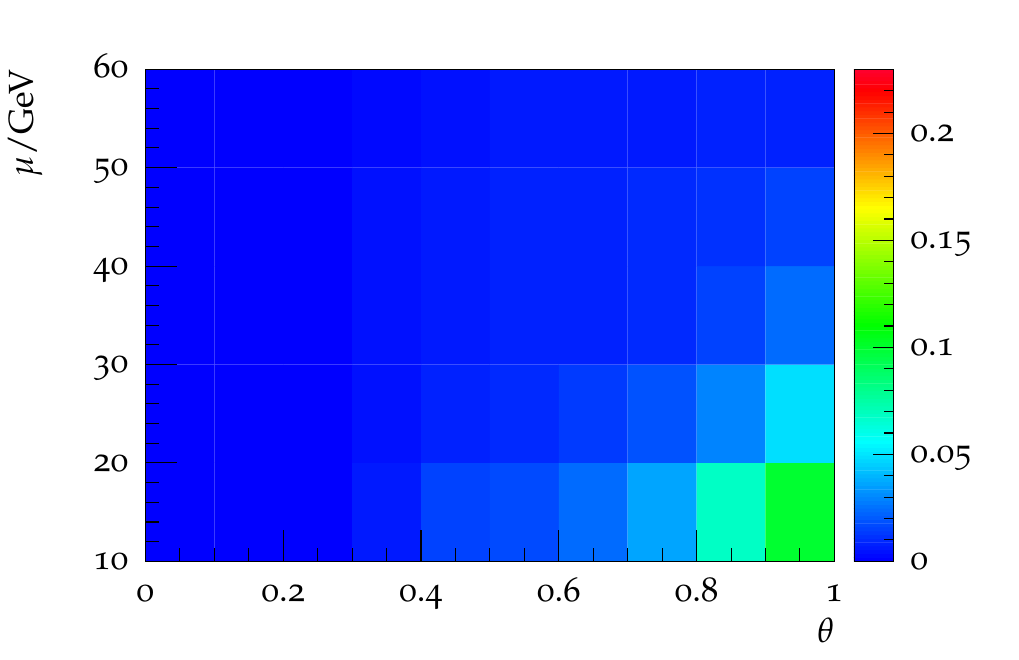}
    \includegraphics[trim=0cm 0cm 1.7cm 0cm, clip=true, height=0.23\textwidth]{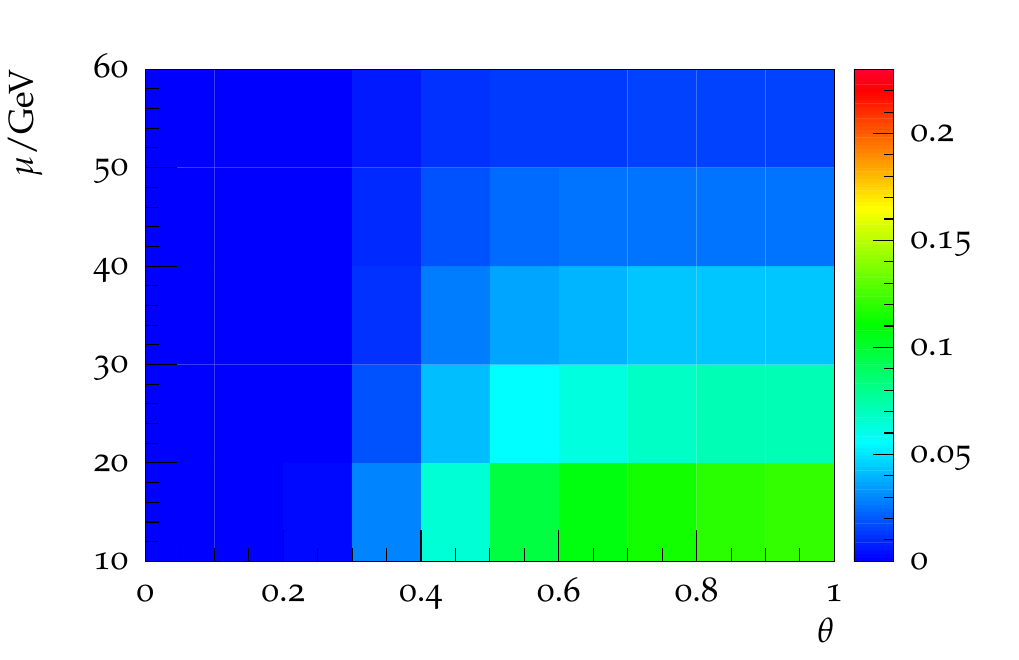}
    \includegraphics[height=0.23\textwidth]{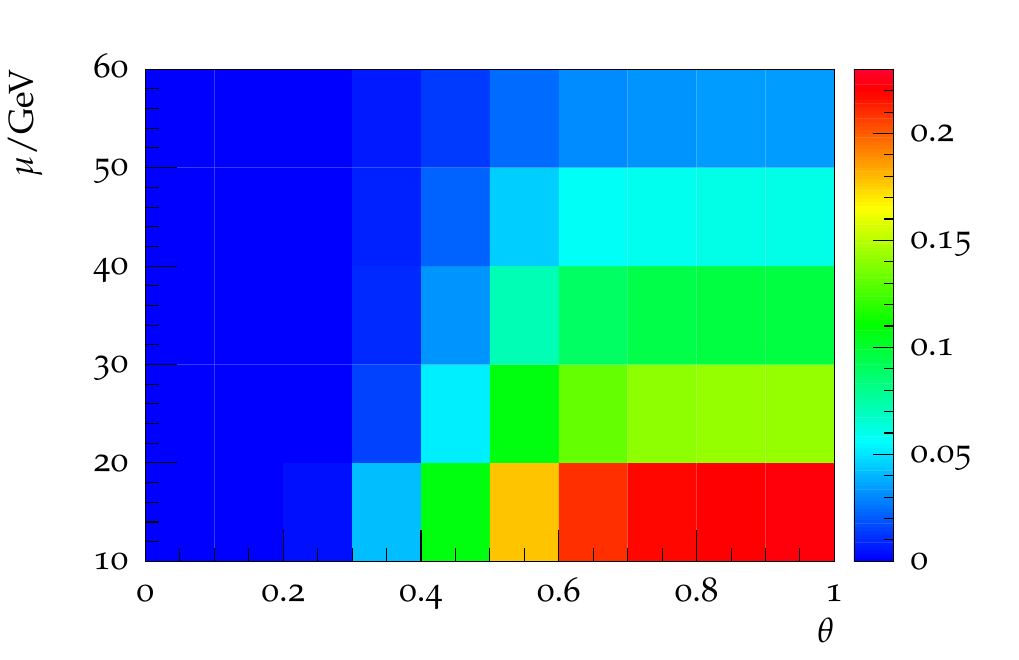}
    
    \begin{minipage}[t]{0.021\textwidth}
    \begin{sideways}$\qquad\qquad \Delta \bar{n}$\end{sideways}
    \end{minipage}
    \includegraphics[trim=0cm 0cm 1.7cm 0cm, clip=true, height=0.23\textwidth]{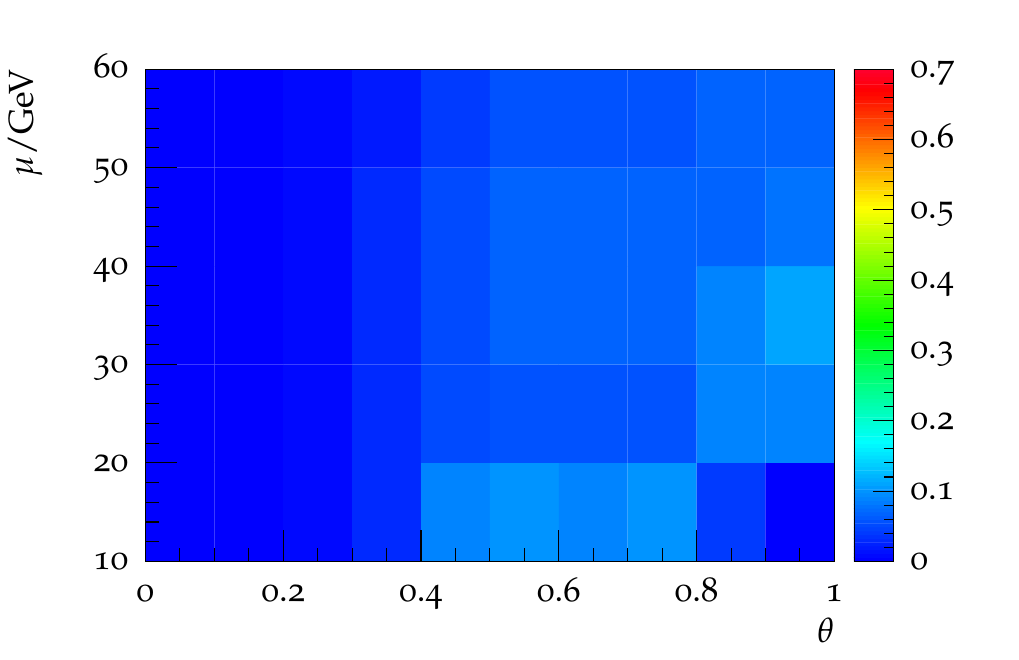}
    \includegraphics[trim=0cm 0cm 1.7cm 0cm, clip=true, height=0.23\textwidth]{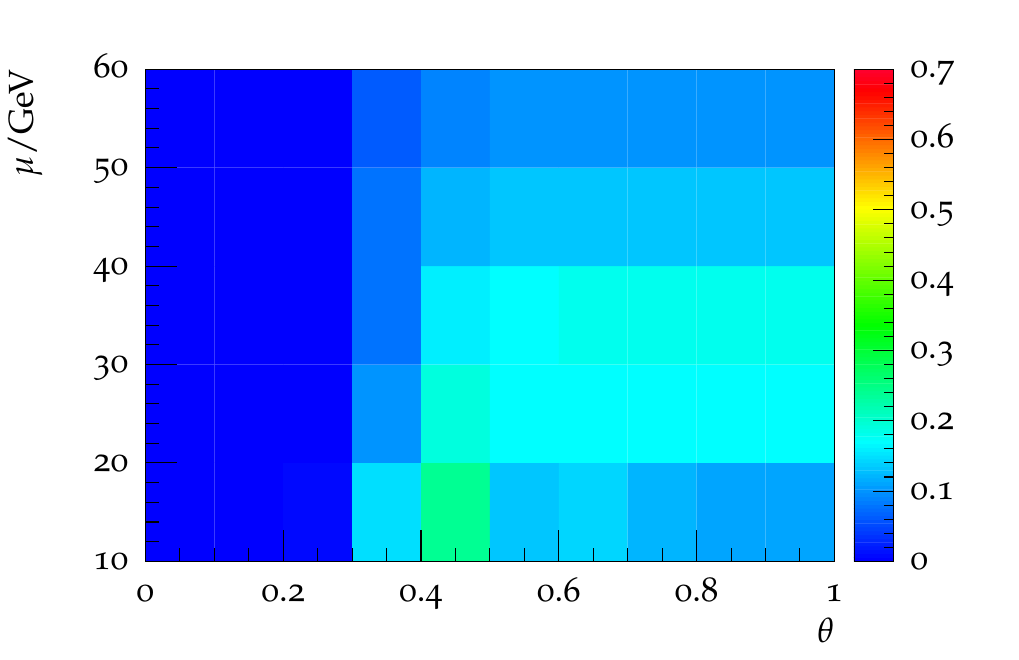}
    \includegraphics[height=0.23\textwidth]{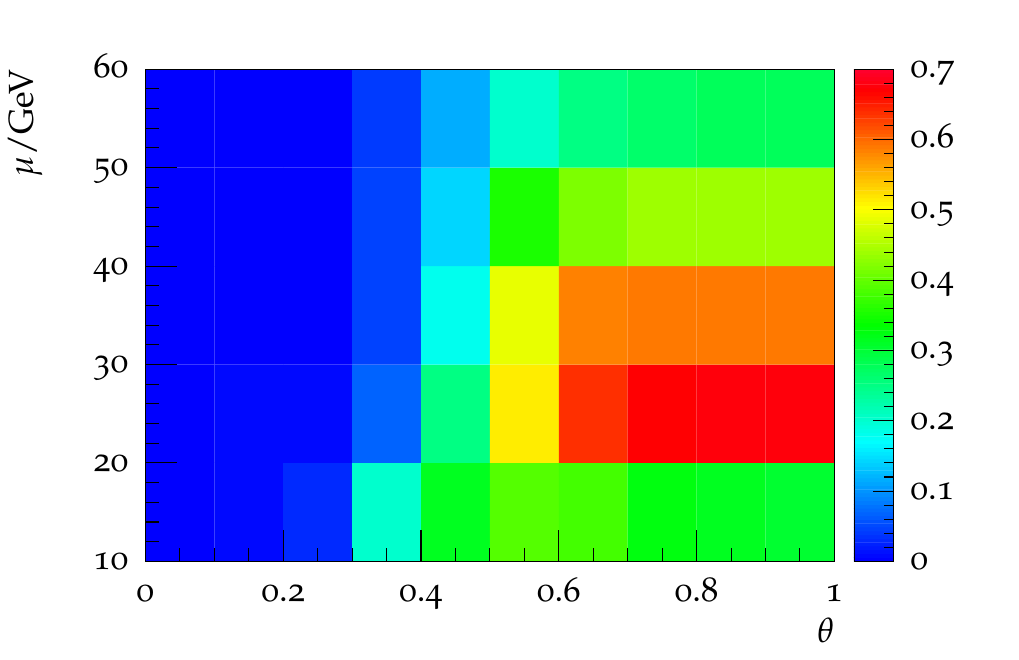}
    
    \caption{Comparison between MJ clustering and its standard counterparts for the anti-$k_T$ (left), C/A (middle) and $k_T$ (right) algorithms. All jets were clustered with $R=0.8$ and $p_\perp\geq 50\GeV$, and only events where $p_\perp^{\text{std}(1)}\geq 150\GeV$ were accepted. The averaged values of the three observables $\Delta R$, $\delta p_\perp$ and $\Delta \bar{n}$ are shown from top to bottom for a range of parameters $\theta$ and $\mu$.}
    \label{fig:sparse}
   \end{center}
  \end{figure}

  The three standard algorithms 
  agree very well in the number of jets $n_\text{std}$, which is 2 (in roughly one in two events) or above.
  We perform a parameter scan for the MJ clustering arguments $\theta$ and $\mu$.
  Figure~\ref{fig:sparse} (bottom panel) shows the difference in the average number of jets per event ($\Delta \bar{n}=\bar{n}_\text{MJ}-\bar{n}_\text{std}$).
  The mutual leading jets (i.e.~the $\min\left[n_\text{MJ},n_\text{std}\right]$ jets with largest $p_\perp$) in each event are matched, and differences between the MJ and standard algorithms are investigated on a jet-by-jet basis. For each pair $(j_\text{MJ},j_\text{std})$, we obtain the $R$-distance ($\Delta R_{j_\text{MJ},j_\text{std}}$) and relative difference in transverse momentum ($\delta p_\perp=\frac{p_\perp^\text{std}-p_\perp^\text{MJ}}{p_\perp^\text{std}+p_\perp^\text{MJ}}$). The upper two panels of figure~\ref{fig:sparse} show the values of these two observables averaged over all matched jet pairs.
  For large parts of the parameter space, the effects of the clustering veto are only limited in this scenario.

  
  Differences between individual jets (upper two rows) are negligibly small in the small-$\theta$ and large-$\mu$ parameter regions for all three jet algorithms. This behaviour is expected as these are the limits where the veto is rendered ineffective.
  The closer the parameters are chosen to the strong-veto region ($\theta\to 1$, $\mu\to 0$), deviations between the vetoed and standard algorithms grow larger. In particular for the $k_T$ algorithm these differences can be substantial, namely $\Delta R\sim 0.5$ and $\delta p_\perp\sim 0.25$ for the considered setup. The C/A and especially the anti-$k_T$ algorithm behave much more moderately under the MJ veto. For the latter, deviations only reach $\Delta R\sim 0.1$ and $\delta p_\perp\sim 0.1$ even in the strong-veto region, and are almost absent in the bulk of parameter space.
  
  Generally the differences between MJ-vetoed and standard clustering are smallest for the anti-$k_T$ algorithm and largest for the $k_T$ algorithm, with the C/A algorithm taking an intermediate position. This characteristic is directly related to the ordering of the cluster sequence, which is crucial in the MJ algorithm. If soft particles are clustered first ($k_T$), it is very likely to induce fake substructure that will fulfill the mass-jump condition at the stage when these soft clusters are recombined. The anti-$k_T$ algorithm on the other hand ignores the parton showering history and clusters around hard prongs. It is therefore much more robust, while the purely angular-based C/A algorithm is moderately prone to vetoing fake soft clusters.
  
  The number of jets is naturally equal or larger in the vetoed algorithms compared to the standard algorithms with equal jet clustering radius (figure~\ref{fig:sparse} lower panels). If however the veto acts too strong, hard jets are split and may not pass the $p_\perp \geq p_\perp^\text{min}$ cut any more, resulting in a decreasing number of jets again. For large minimum jet transverse momentum close to $p_\perp^{\text{std}(1)}$, say $p_\perp^\text{min}=100\GeV$ for our analysis, $\Delta\bar{n}$ ultimately becomes negative.
  
  Also for other jet clustering radii and $p_\perp$ thresholds, results are qualitatively very similar to the ones described above, so we omit further plots.


\subsection{Busy environment: boosted top quarks}

  Tagging boosted top quarks is an important target in many current experimental studies and also an ideal playground to investigate the performance of MJ clustering in busy environments.
  In order to probe the moderately boosted energy regime and illustrate the algorithm, we simulate top pair production via a hypothetical heavy vector boson,
  \begin{align}
   p p \to Z' \to t\bar{t} \to \text{hadrons}
  \end{align}
  for three different resonance masses $m_{Z'}=500$, $700\GeV$, and $1\TeV$. 
  The first sample results in fat jets (Cambridge/Aachen with $R=1.5$, $p_\perp\geq 200\GeV$) whose $p_\perp$ distribution drops steeply to mimic top quarks produced in SM processes. The latter two samples emulate a generic heavy resonance and yield top quarks with transverse momentum peaking around $\sim300$ and $\sim450\GeV$, respectively.
  Those fat jets are fed to the HEPTopTagger~\cite{Plehn:2010st}, which performs the following three-step procedure.
  \begin{enumerate}
   \item Subjets are obtained from the fat jet via mass-drop unclustering as outlined in section~\ref{sec:algorithm:md}.
   \item A filtering stage~\cite{Butterworth:2008iy} is applied to reduce QCD effects:
   the constituents of three subjets are reclustered with a smaller radius
   $R_\text{filter}=\min\left( 0.3, \Delta R_{ij}\right)$. The new top candidate subjets are then formed by reclustering the $n_\text{filter}=5$ hardest small jets to exactly three jets. This constitutes a possible top candidate if the combined mass lies within $m_t\pm 25\GeV$.\\
   In the case that more than three subjets were found in the first step, only the three-subjet combination with a filtered mass closest to the real top mass is considered.
   \item Cuts on subjet mass ratios ($m_{12}$, $m_{13}$, $m_{23}$ calculated from the $p_\perp$-ordered subjets~\cite{Plehn:2010st}) determine whether or not the candidate is tagged as top; in addition, the candidate's transverse momentum is required to be $\geq 200\GeV$.
  \end{enumerate}
  
\begin{figure}[t]
   \begin{center}
   \begin{minipage}{0.02\textwidth}
   \end{minipage}
   \begin{minipage}{0.48\textwidth}
    \centering
    HEPTopTagger with MD un-clustering
   \end{minipage}
   \begin{minipage}{0.48\textwidth}
    \centering
    HEPTopTagger with MJ clustering
   \end{minipage}

   \begin{minipage}[t]{0.02\textwidth}
   \begin{sideways}$\qquad\quad m_{Z'}=500\GeV$\end{sideways}
   \end{minipage}
   \includegraphics[width=0.45\textwidth]{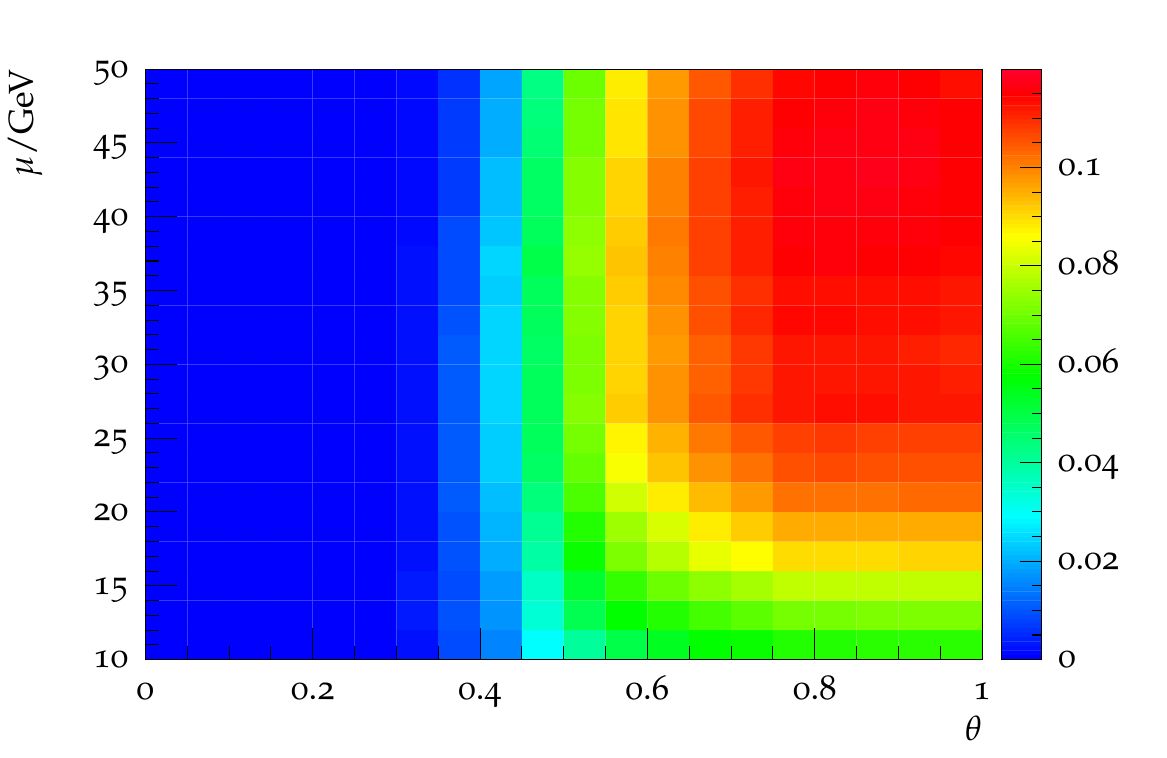}
   \includegraphics[width=0.45\textwidth]{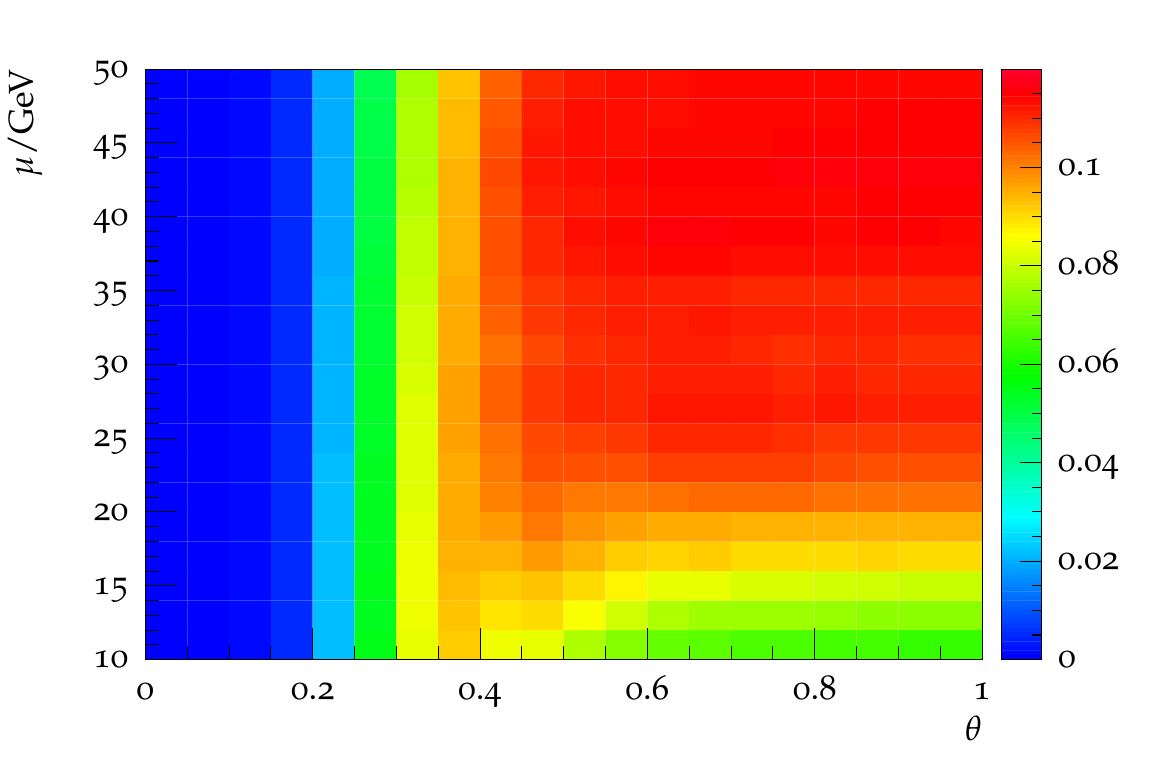}
   
   \vspace{-0.4cm}
   \begin{minipage}[t]{0.02\textwidth}
   \begin{sideways}$\qquad\quad m_{Z'}=700\GeV$\end{sideways}
   \end{minipage}
   \includegraphics[width=0.45\textwidth]{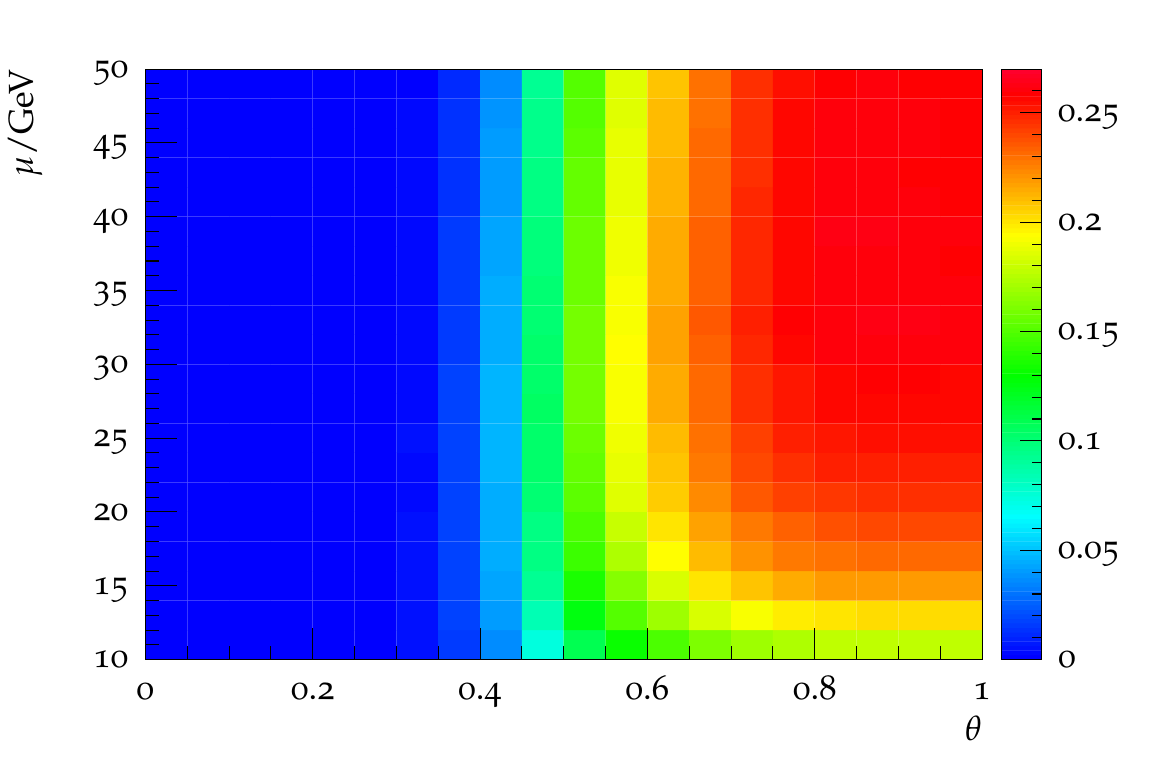}
   \includegraphics[width=0.45\textwidth]{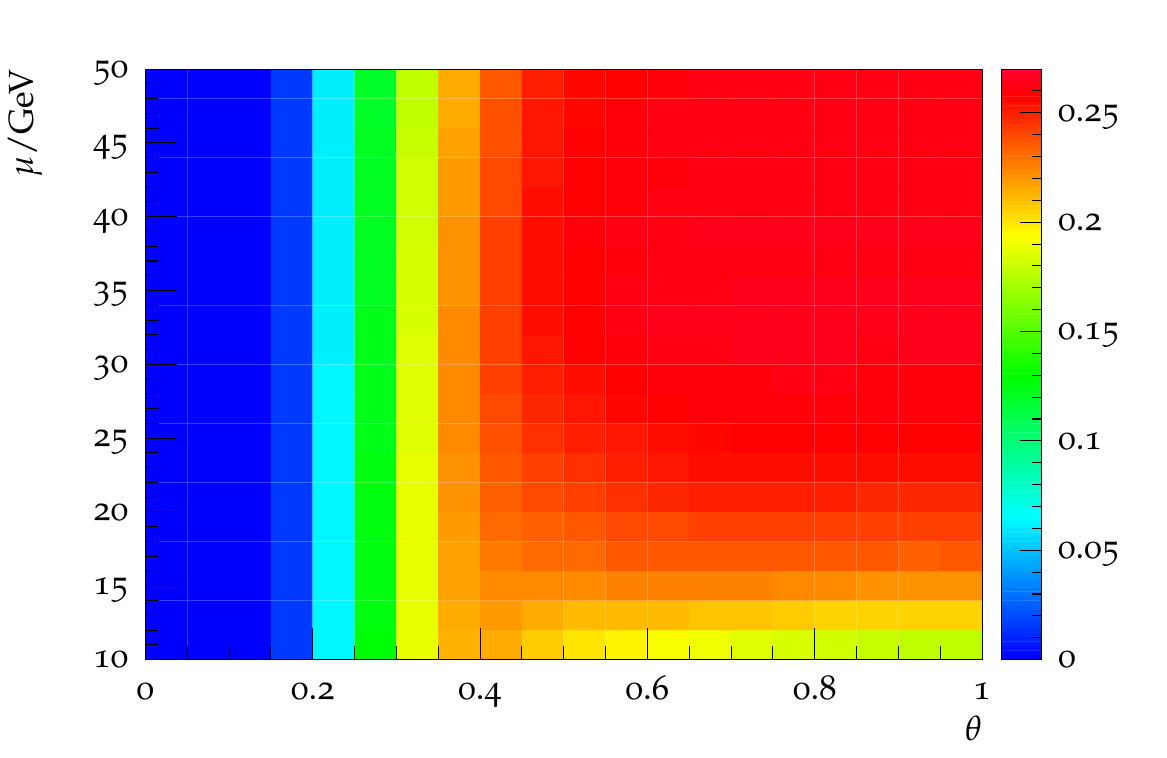}
   
   \vspace{-0.4cm}
   \begin{minipage}[t]{0.02\textwidth}
   \begin{sideways}$\qquad\quad m_{Z'}=1\TeV$\end{sideways}
   \end{minipage}
   \includegraphics[width=0.45\textwidth]{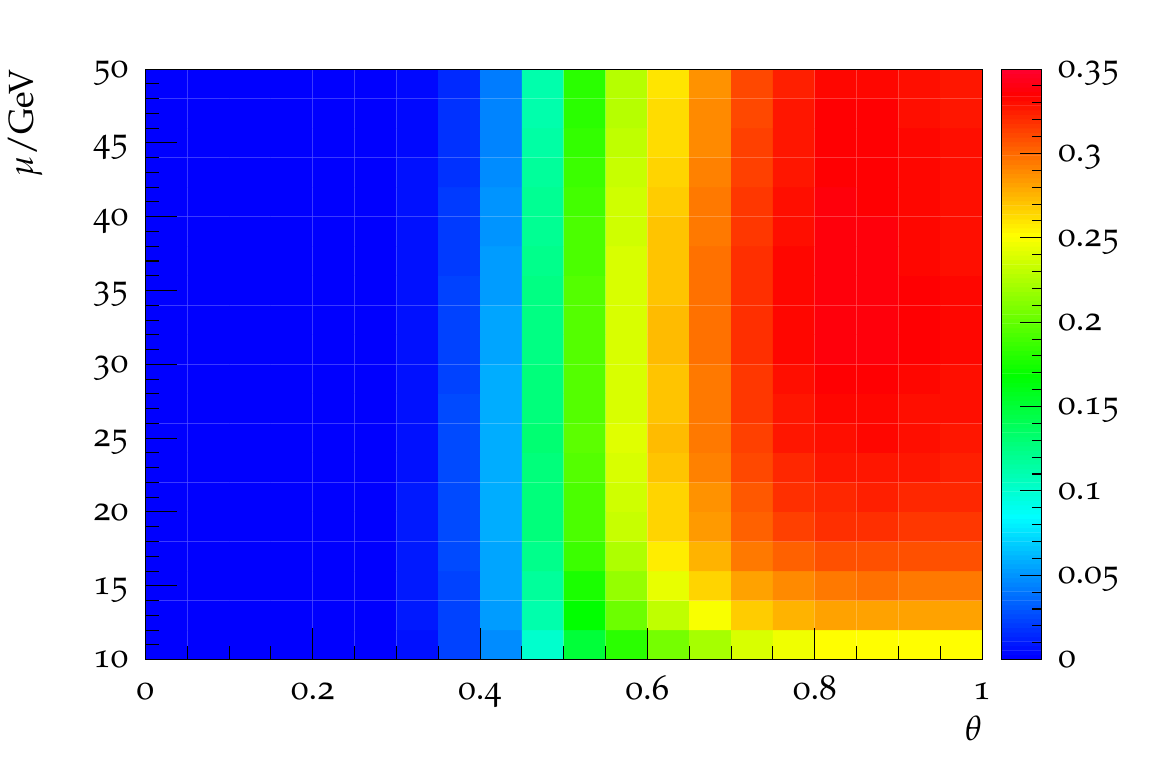}
   \includegraphics[width=0.45\textwidth]{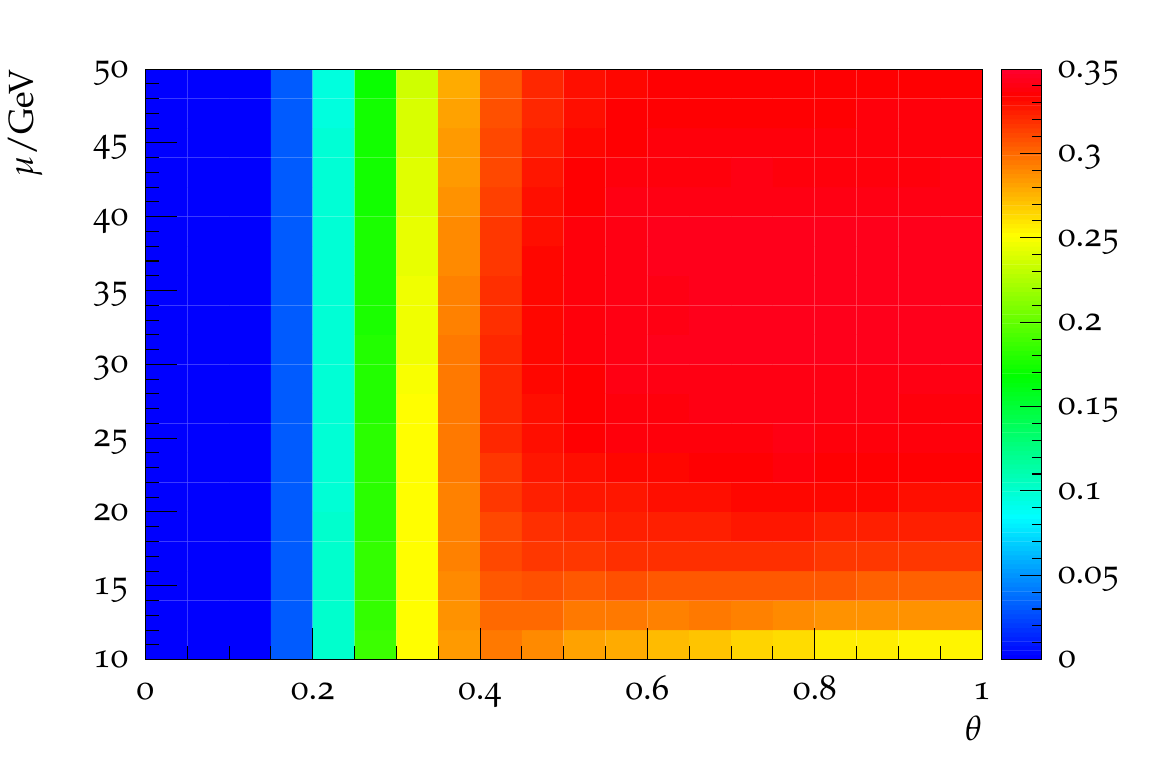}
   
   \caption{Top tagging efficiency $\epsilon$ for the HEPTopTagger with MD un-clustering (left) and MJ clustering (right). For both algorithms the parameter space $\theta,\mu$ is scanned. 
   From top to bottom, the panels show signal rates for the $m_{Z'}=500\GeV$, $700\GeV$, and $1\TeV$ samples.
   }
   \label{fig:heptopjump2d}
   \end{center}
  \end{figure}

  For comparison with our veto algorithm, we apply the same HEPTopTagger algorithm but where the subjets are now obtained directly with MJ clustering, starting from the fat jet's constituent particles. Steps 2 and 3 remain unchanged such that the difference in tagging performance can be directly compared.
  We take $R=\infty$ and scan the parameter space in $\theta$ and $\mu$. Results are based on each 10,000 signal and background events (QCD dijets with $\hat{p}_\perp^\text{min}=150\GeV$) generated with \pythia8 and analyzed within \rivet. The resulting tagging efficiencies $\epsilon=\frac{\#\text{tags}}{\#\text{fat jets}}$ are shown in figure~\ref{fig:heptopjump2d}.\footnote{Fat jets that deviate too much from their Monte Carlo truth top quark ($\Delta R_{j^\text{fat},t^\text{MC}} > 0.6$) are ignored in signal events.}
  
  
  Indeed the peak tagging efficiencies are equal for both algorithms and constant over a relatively large part of parameter space. However, as argued in section~\ref{sec:algorithm:properties}, MJ jet finding allows for well-performing top tagging in a much wider range in the parameter $\theta$. The reason for this behaviour flies in the absence of an equivalent to the cascade mass drops experienced in MDT's (such as $t\to bW^+\to bjj'$).
  This feature can also be directly seen in figure~\ref{fig:heptopjump2d} where in the MDT case (left) the onset of top tagging is around $\theta=0.5\approx\frac{m_W}{m_t}$, whereas for MJ clustering (right) the characteristic scale is much lower.
  In particular, lower values of $\theta$ correspond to a much stricter identification of separate jets, which might turn out beneficial for background rejection.
  
  The observed overall increase in tagging efficiency for larger resonance masses $m_{Z'}$ is a simple consequence of the underlying kinematics. The majority of fat jets carry a larger transverse momentum than the respective initiating top quark. As a result, the very last cut ($p_\perp^\text{top candidate}\geq 200\GeV$) rejects many moderately-boosted candidates even in the case of perfect reconstruction. With larger boost (corresponding to larger $m_{Z'}$), this fraction becomes smaller.
  
\begin{figure}[t]
  \begin{center}
   \begin{minipage}{0.1\textwidth}
   \end{minipage}
   \begin{minipage}{0.31\textwidth}
    \centering
    $m_{Z'}=500\GeV$
   \end{minipage}
   \begin{minipage}{0.35\textwidth}
    \centering
    $m_{Z'}=700\GeV$
   \end{minipage}
   \begin{minipage}{0.22\textwidth}
    \centering
    $m_{Z'}=1\TeV$
   \end{minipage}

   \begin{minipage}[t]{0.02\textwidth}
   \begin{sideways}$\qquad$ hadron level\end{sideways}
   \end{minipage}
   \includegraphics[width=0.31\textwidth, trim=5cm 14.5cm 4cm 4.5cm, clip=true]{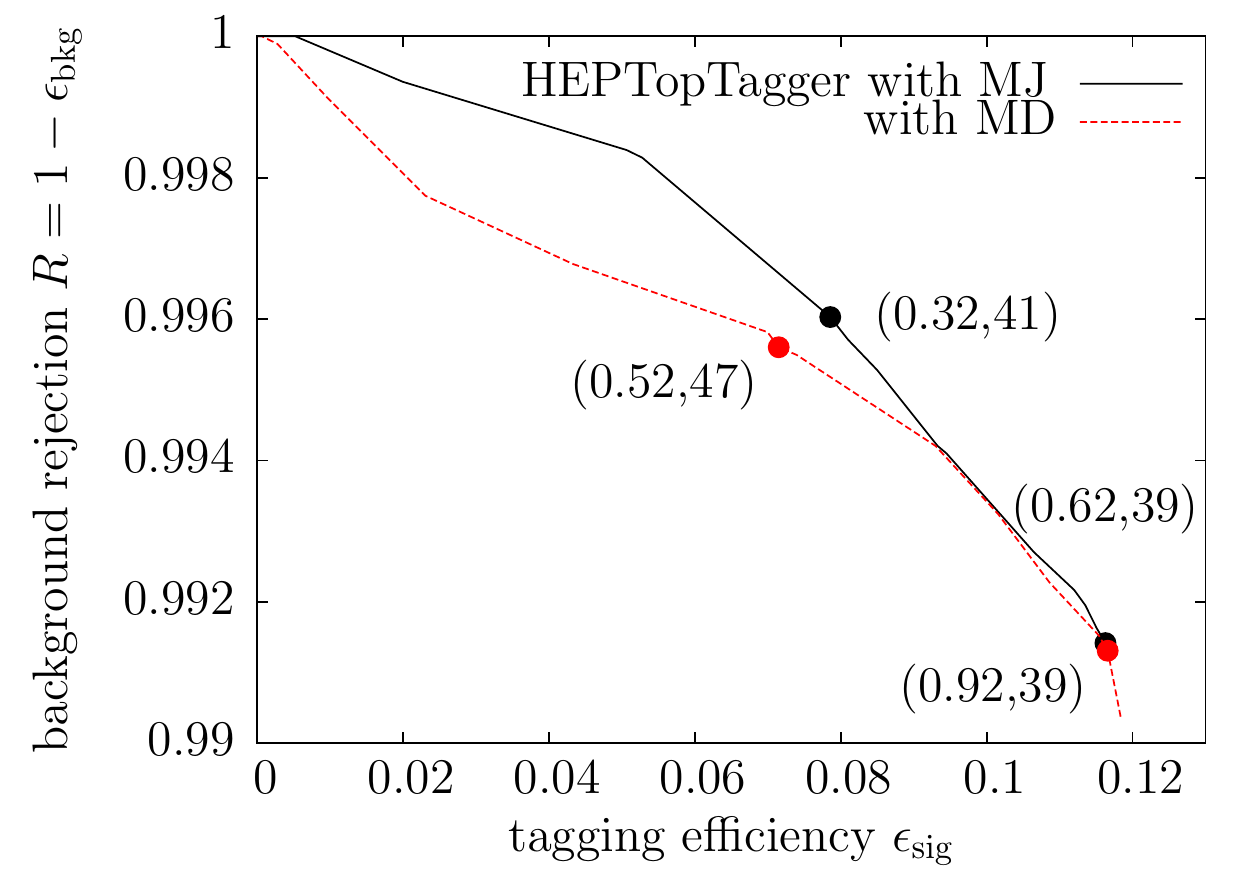}
   \includegraphics[width=0.31\textwidth, trim=5cm 14.5cm 4cm 4.5cm, clip=true]{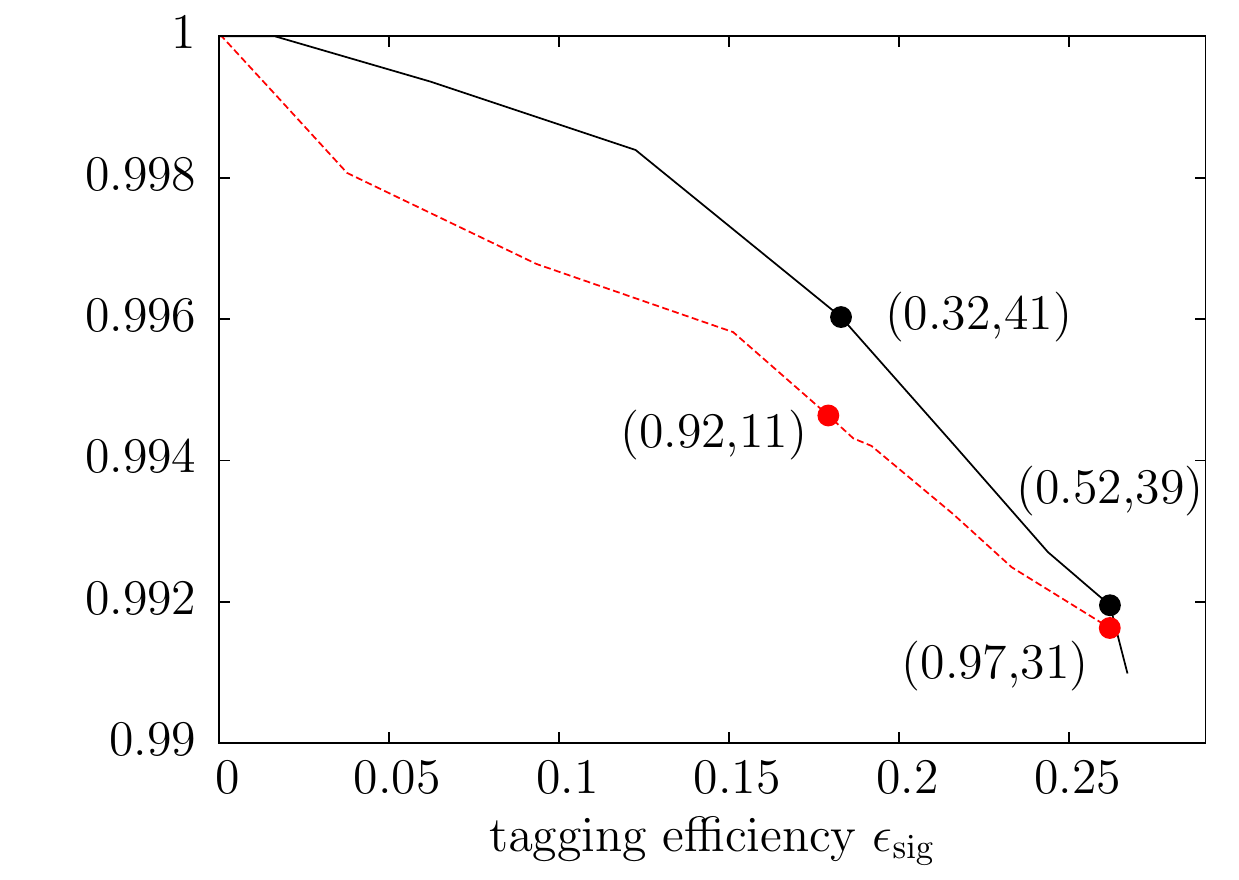}
   \includegraphics[width=0.31\textwidth, trim=5cm 14.5cm 4cm 4.5cm, clip=true]{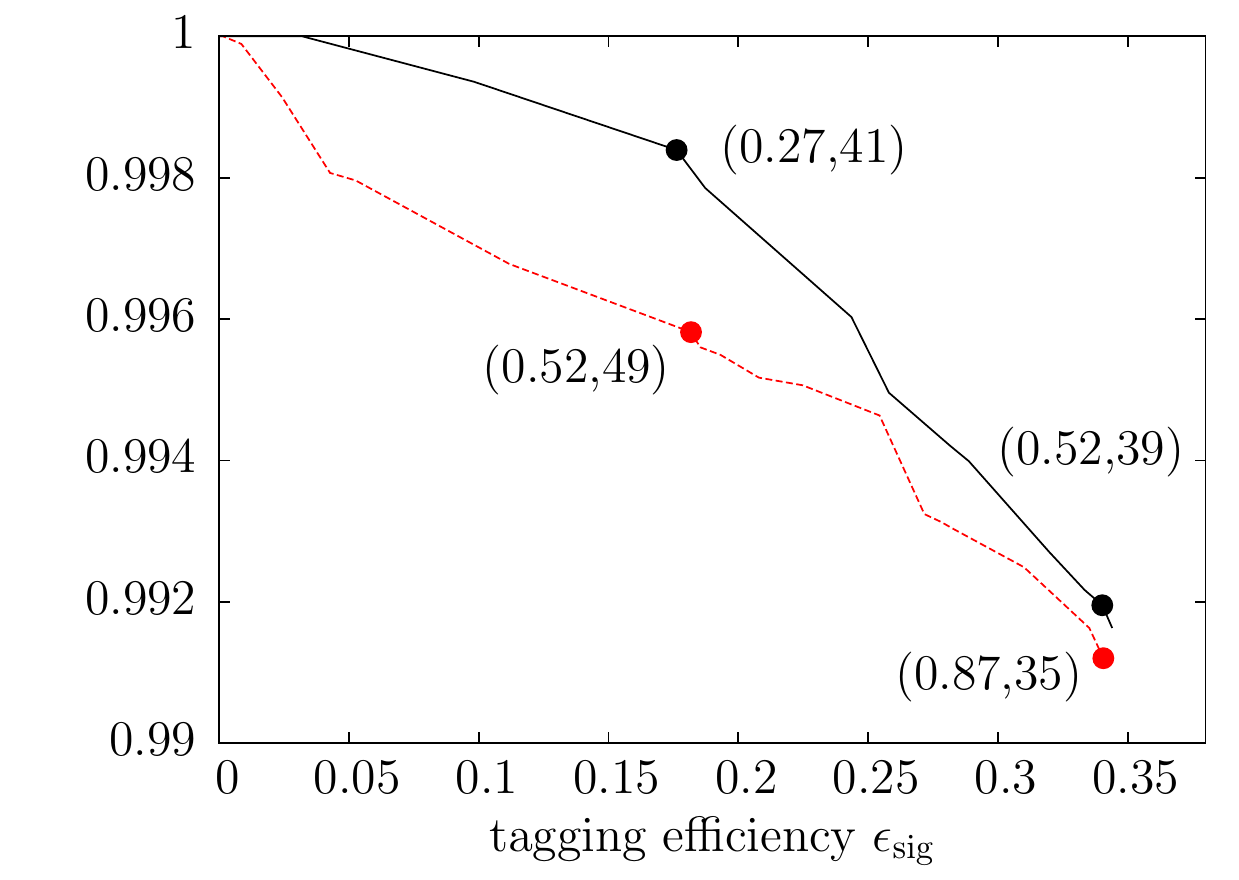}
   
   \begin{minipage}[t]{0.02\textwidth}
   \begin{sideways}$\qquad$ clustered level\end{sideways}
   \end{minipage}
   \includegraphics[width=0.31\textwidth, trim=5cm 14.5cm 4cm 4.5cm, clip=true]{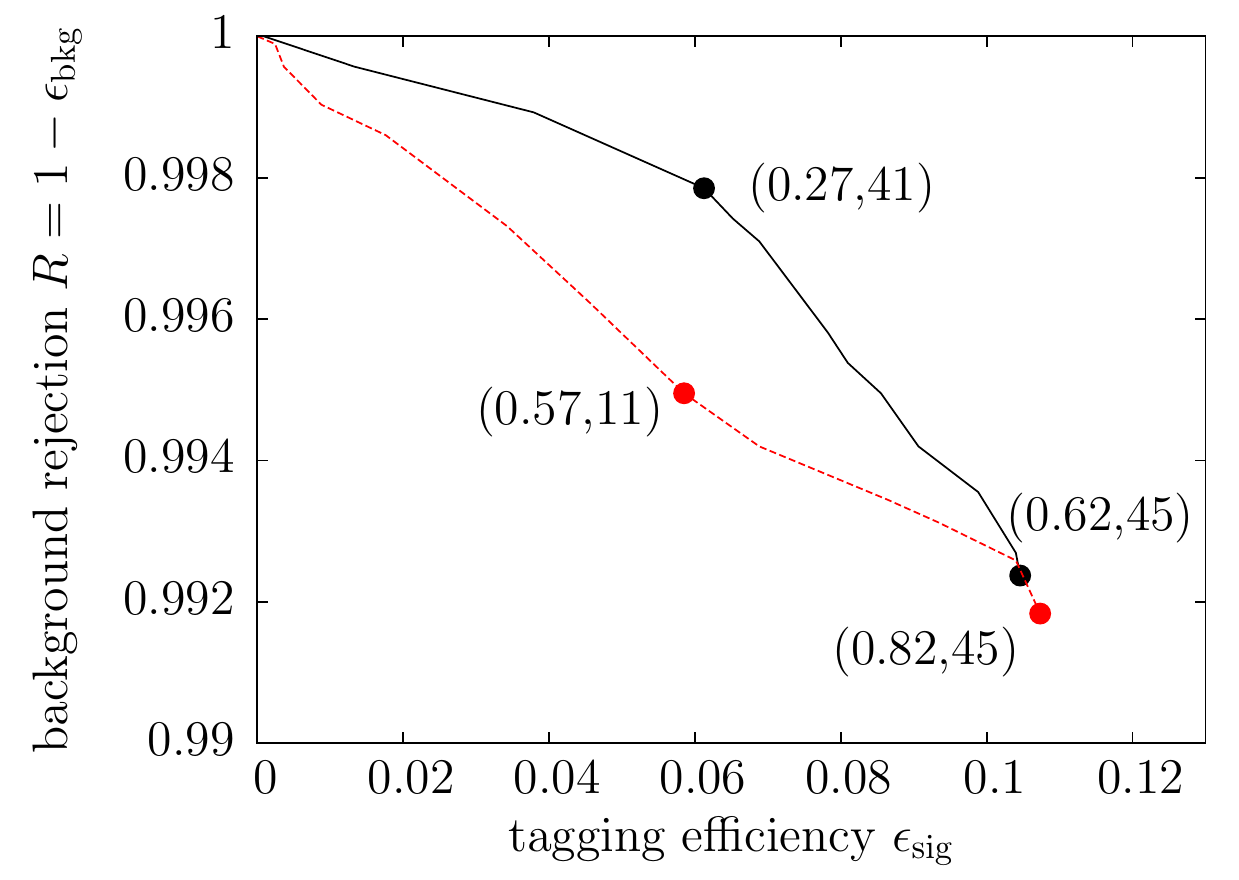}
   \includegraphics[width=0.31\textwidth, trim=5cm 14.5cm 4cm 4.5cm, clip=true]{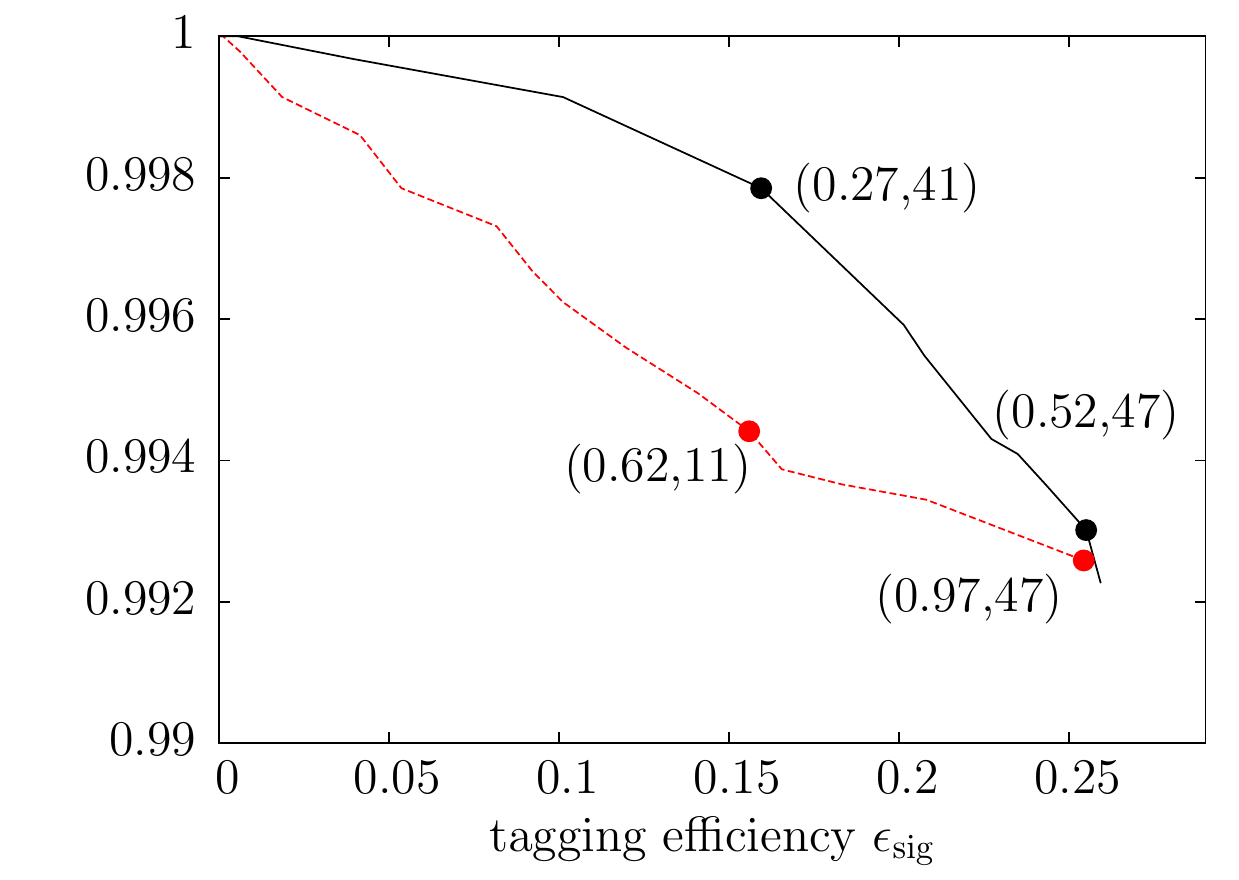}
   \includegraphics[width=0.31\textwidth, trim=5cm 14.5cm 4cm 4.5cm, clip=true]{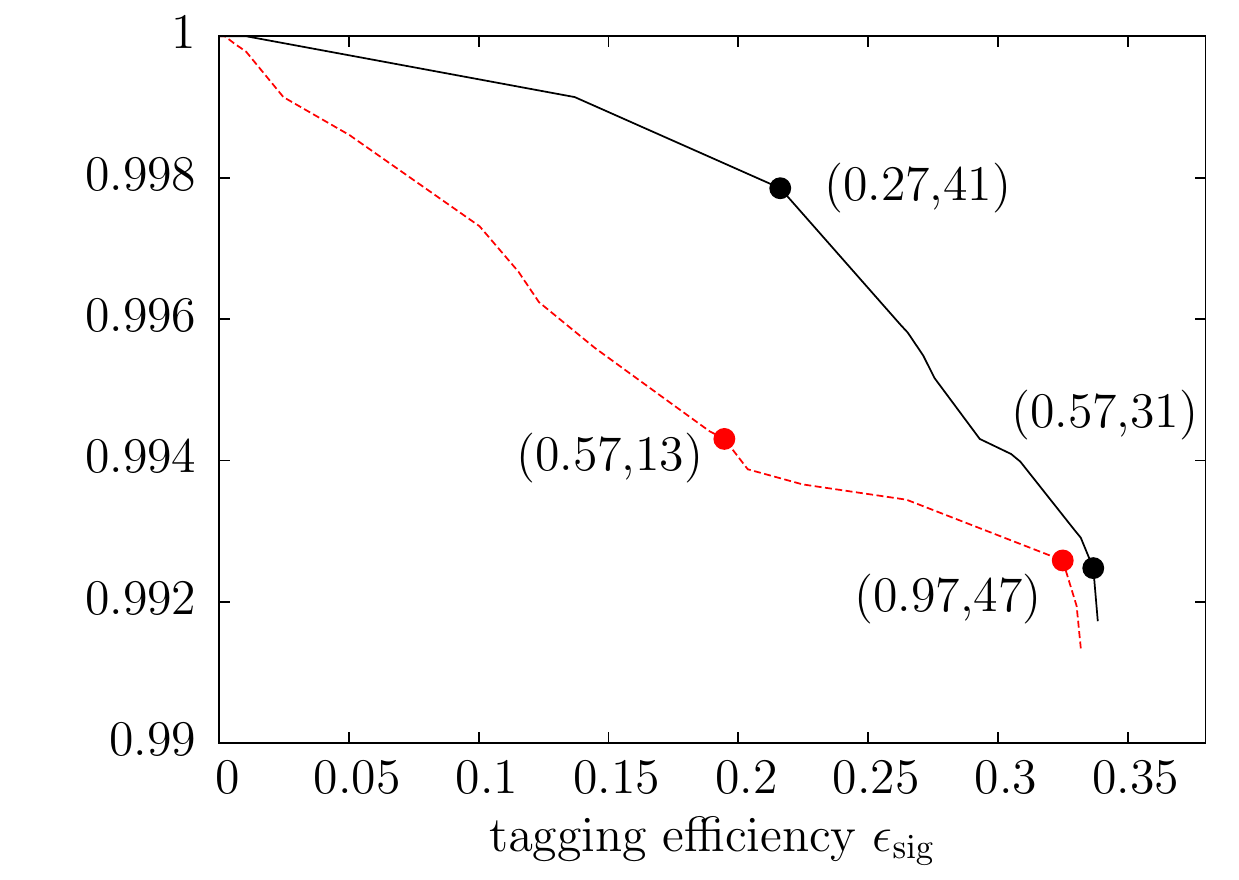}
   
   \caption{Receiver-operating characteristic (ROC) curves for top tagging using the HEPTopTagger. Subjet finding with the MJ clustering algorithm (black solid) is compared to the original algorithm, which employs MD un-clustering (red dashed). From left to right, the upper panels show results at hadron level for the $m_{Z'}=500\GeV$, $700\GeV$ and $1\TeV$ sample, respectively.
   The lower panels are similar but obtained from hadrons centred into (0.1, 0.1) cells in $\eta$--$\phi$ space to emulate finite detector resolution. 
   Parameters at exemplary benchmark points are given for illustration in the format $(\theta,\mu/\text{GeV})$. Note that different parameter points can yield similar efficiencies, and that the benchmark points are chosen somewhat arbitrarily in this sense.
   If high purity is desired, MJ clustering gives improved performance.}
   \label{fig:ROC_heptoptag}
  \end{center}
  \end{figure}

  
  Figure~\ref{fig:ROC_heptoptag} compares the receiver-operating characteristic (ROC) curves of the original HEPTopTagger and the modified algorithm where MD unclustering has been replaced by MJ clustering.\footnote{These curves are obtained from the full parameter scan. Among all setups $(\theta,\mu)$ that give a similar signal tagging efficiency, only the one that yields the highest background rejection is picked and plotted.}
  It is observed that signal tagging efficiency and background rejection coincide for large efficiencies, giving 
  $\epsilon_\text{sig}\approx 0.12$ and $R=1-\epsilon_\text{bkg}\approx 0.991$ for the $m_{Z'}=500\GeV$ sample,
  $(0.26,0.991)$ for the $m_{Z'}=700\GeV$ sample,
  and $(0.34,0.992)$ for the $m_{Z'}=1\TeV$ sample, respectively. 
  These values correspond to the plateau at large $\theta$ and medium-to-large $\mu$ in figure~\ref{fig:heptopjump2d}.
  However due to the enlarged parameter space, the MJ algorithm outperforms the standard procedure and should be preferred in the transition (high-purity) region. This result is even more pronounced if limited detector resolution is taken into account. For our simple analysis, this is implemented by applying a cellular grid in the $\eta$--$\phi$ plane and replacing all stable hadrons to the centre of their respective cells. 
  For most working points, the inevitable decrease in performance is less pronounced when MJ clustering is used. At maximum tagging efficiencies the two algorithms still give the same results.

\section{Conclusions}
\label{sec:conclusions}

  We developed and investigated a new jet clustering algorithm that includes a recombination veto based on jet mass. In this mass-jump (MJ) procedure, the clustering radius $R$ now acts as an upper limit on jet size and the merging of two hard prongs is prevented. We showed that in sparse events with well-separated jets, the effect of the veto is very limited in a large range of the parameter space. Also the anti-$k_T$ clustering algorithm is more robust against fake two-prong substructure than the Cambridge/Aachen and $k_T$ algorithms.
  In the dense environment of hadronically decaying boosted top quarks, MJ clustering gives results comparable to those of mass-drop taggers (MDT) by which the veto was inspired in the first place; the main difference being that cascade mass drops as present in MDT's are avoided, which in turn allows for stricter threshold parameters. The larger parameter space then leads to improved ROC curves 
  for the HEPTopTagger when the mass-drop procedure is replaced by MJ clustering.
  
  Until the veto is interposed, MJ jet clustering proceeds identically to its standard counterpart.
  In particular, no soft radiation is removed and after the veto (multiple vetoes) additional jets are formed from the remaining particles.
  Especially in realistic scenarios when soft QCD radiation (from underlying event or pile-up) is present, the application of grooming techniques can improve jet shape observables by removing this uncorrelated energy.
  
  Jet algorithms with a terminating veto
  are a promising tool for collider experiments as they make room for more flexibility. The optimal clustering radius depends on various parameters such as the type of initiating particle, its energy or transverse momentum, and the surrounding topology of the event. The MJ veto automatically adjusts the jet radius such that hard substructure is separated into isolated jets. This feature may prove helpful in a variety of events where jets are not well-separated.

\begin{acknowledgments}
The author is grateful to Koichi Hamaguchi for helpful discussion and also Go Mishima for comments on the manuscript.
This work was supported by the Program for Leading Graduate Schools.
\end{acknowledgments}



\providecommand{\href}[2]{#2}\begingroup\raggedright\endgroup


\end{document}